\documentstyle{article}
\begin{document}
\hspace{1cm}  \vspace{1cm}

\begin{center}
	 {\Large {\bf 	Nonlinear Regimes in Thermostats
			of Berendsen's Type \vspace{7mm} \\
		 }
		 V.L. Golo$^{\dagger}$  and
		 K.V. Shaitan$^{\dagger \dagger}$,
		 \vspace{2cm}\\
	 }
	   $^{\dagger}$ Department of
	   Mechanics and Mathematics \vspace{1mm} \\
	   $^{\dagger \dagger}$Department of   Biology  \\
	   Moscow University, Moscow 119899  \\
	   RUSSIA \vspace{5mm} \\
\end{center}

\noindent
{\Large {\bf Abstract} } \vspace{5mm}

\noindent
We consider the models of relaxational dynamics within the
framework of  Berendsen's and Nos\'e---Hoover's thermostats.
On studying the crucial case of ideal gas we come to the
conclusion that both models mentioned above do not allow for
describing the true thermodynamical equilibrium.

\pagebreak

\noindent
{\bf 1. Introduction}

\noindent
The thermodynamical concept of thermostat supposes
that a system under consideration is a small part of a large system, or
thermostat, which is big enough so as to be insensitive to changes
inside the small one, and thus a very large  number of particles
is generally required to the effect. On the other hand,
the small system may be in its turn rather tiny, even comprise only
one particle.
The thermostat's being  macroscopical, creates serious difficulties
for numerical modeling thermodynamical systems at fixed temperature and
therefore, there is a need for simplified models that could
accommodate the requirement of fixed temperature.  The most straightforward
approach to the problem is the use of  the so-called
collision thermostat, that provides adequate means for describing the
system's interaction with ambient media.  This method has its drawbacks in
that it is time consuming and puts severe constraints on the
precision level. Hence, it is very tentative to find certain simpler
ways of modeling, using systems that do not involve
extremely large number of particles or stochastic
equations.  The most familiar models in this respect are
the Berendsen, \cite{ber} and the Nos\'e---Hoover, \cite{nos}, one.
The idea of both is the same: we shall introduce
in the right hand sides of a system under consideration relaxation
terms that contain a dissipation parameter that should mimic the control
exercised on the system by thermostat. To this end one uses a relation,
of the most simple form possible, that requires that the kinetic energy
of the {\it total} system should be close to that prescribed by the heat
energy determined by thermostat parameters. Thus, from
the mathematical point of view,  equations obtained in this way
are an extension of the initial ones through the introduction of nonlinear
dissipative terms in the right hand side.
The prescription is quite general,
appealing intuitively, and easy to implement.
But it reduces the problem to that of a finite nonlinear dissipative system,
with no macroscopical features left and
essential thermodynamical properties being muted.
Cosequently, one may wonder as to whether
the physical nature of the phenomenon has been preserved.

\noindent
In this paper we are going to consider examples of applying
the prescription to a simple, but extremely important, case of ideal gas.
Its analysis brings us to the conclusion that
the Berendsen and Nos\'e --- Hoover systems allow for
interesting nonlinear regimes, even for this simple case,
but do not generate the system's thermodynamical equilibrium.

\noindent
{\bf 2. Mechanical model}

\noindent
Let us recall the construction of the  thermostats mentioned above.
We shall begin with the Berendsen one.
Suppose we have a system of particles whose dynamics is governed by
the Newton equations

$$
 m_i \ddot{ \vec r_i } = - \frac{\partial}{\partial \vec r_i} U(r)
$$

\noindent
The presence of a thermostat at temperature $T$ is claimed to be
accommodated through the introduction of non-linear dissipative terms in
the equations indicated above, that is

\begin{equation}
	 m_i \ddot{ \vec r_i}  =
	- \frac{\partial}{\partial \vec r_i} U(r)
	- \gamma \dot{\vec r_i}
	\label{ber}
\end{equation}

\noindent
with the dissipative coefficient $\gamma$ being of the form

\begin{equation}
 \gamma = \alpha \left[	\frac{1}{3 k_b T N}
		       \sum_{i=1}^N m_i \dot{ \vec r_i}^2
			- 1 \right]
		       \label{berkof}
\end{equation}

\noindent
where $N$ is the number of particles. Intuitively, it is "clear" that
the model described above  by equations (\ref{ber}), (\ref{berkof})
should guarantee that the system be at temperature T, or at least its
total kinetic energy be close to that value. The distribution of the
kinetic energy per particle is not so clear.
To be more precise equation (\ref{ber})  in the notations of
paper \cite{ber} contains
the kinetic energy in the denominator;
but it is not important for what follows.

\noindent
In the Nos\'e --- Hoover thermostat the dissipative variable $\gamma$ is
treated on equal footing with other dynamical variables, and instead
of the algebraic equation (\ref{berkof}) they write the differential
equation

\begin{equation}
 \dot{ \gamma} =  \alpha \left[
			\frac{1}{3 k_b T N}
			\sum_{i=1}^N m_i \dot{ \vec r_i}^2 - 1
			\right]  \label{nos}
\end{equation}

\noindent
The system of equations (\ref{ber}) and (\ref{nos})  is also quite
appealing intuitively and looks excellently physically motivated.
Again, at first sight it seems that the kinetic energy should be
close to the heat one, at least if the system is kept long enough
in the regime, and again one may have certain doubts, from the very
beginning, as to its distribution with respect to degrees of freedom.
\vspace{1cm}

\noindent
{\bf 2. Ideal gas}

\noindent
Let us consider an ensemble of $N$ noninteracting identical particles of mass
$$
  m_i = m, \quad i = 1,2, \ldots , N
$$
that is an ideal gas, in a thermostat. Consider first the Berendsen one.
Then the equation \ref{ber} takes the form

\begin{equation}
	 m \ddot{ \vec r_i}  =
	- \gamma \dot{ \vec r_i}
	\label{ber1}
\end{equation}

\noindent
in which $i$ takes the values $1, 2, \ldots ,N $. It is more convenient
to use the momenta
$$
   \vec p_i = m \dot{ \vec r_i}, \quad i = 1,2, \ldots , N
$$
and cast equations (\ref{ber}) in  the form

\begin{equation}
	  \dot{ \vec p_i}  = - \gamma  \vec p_i
	\label{ber1p}
\end{equation}

\noindent
On multiplying equations in (\ref{ber1p}) with $\vec p_i, i = 1,2, \ldots, N$
and next taking their sum, we obtain the equation for
the energy $E$

\begin{equation}
	 \frac{d}{dt} E  = - 2 \alpha E
	 \left( \frac{2 E}{3 k_b T N}  -1 \right)
	 \label{ber2}
\end{equation}

\noindent
In equilibrium the value of energy is fixed, and
therefore, the right hand side
of the equation given above be equal to zero, so that  we obtain
the two values: (1) $E = 0$, which is meaningless, and
(2) $ E = \frac{3}{2} k_b T N$,
which is the right one. But the second requirement still remains, that is
the existence of Maxwell's distribution for velocities of particles.
Here we run across serious difficulties. The main point is that
thermostat ought to bring the system to the equilibrium,
even if its initial configuration is outside the latter. Let us consider an
initial state of the gas that comprises two beams of particles,
the first one comprising  of $N_1$ particles,
all of them having the same momentum
$\vec P_1$, and the second one of the remaining $N_2 = N - N_1$ particles,
all having the momentum $\vec P_2, \quad \vec P_1 \neq \vec P_2$. It is
to be noted that equations (\ref{ber1p}) and (\ref{berkof})  are invariant
with respect to permutations of the indices $i$, that is we may interchange
the particles so that solutions to equations  (\ref{ber1p}), (\ref{berkof}),
remain valid. This is a form of Gibbs' principle, that is  particles of
ideal gas being not distinguishable from each other. But the particles of
the first beam have the same initial momenta, $\vec P_1$, and therefore
they are described by the same equation in which
$$
  \vec p_i  = \vec P_1, \quad i = 1,2, \ldots , N_1
$$
and the same is true for the particles of the second beam, in which
$$
  \vec p_i  = \vec P_2, \quad i = N_1 +1,N_1+2, \ldots , N
$$
Thus, we may write only two equations for the momenta $\vec P_1, \, \vec P_2$
instead of $N$ equations (\ref{ber1p}), that is

\begin{equation}
 \dot{\vec P_{\nu}} = - \alpha
			   \left[ \frac{N_1 P_1^2 + N_2 P_2^2}{3 m k_b T N}
				- 1
			   \right ] \vec P_{\nu}, \quad \nu = 1,2
			       \label{redber}
\end{equation}

\noindent
From the last equation we infer that the locus of stationary states for a gas
having an initial state consisting of two beams of $N_1, N_2$
of particles, respectfully, is the ellipse in plane $ P_1, P_2 $,
given by the equation

\begin{equation}
  \frac{N_1}{3 k_b T N} P_1^2 + \frac{N_2}{3 m k_b T N} P_2^2 = 1
  \label{ellipse}
\end{equation}

\noindent
In Fig.1 we illustrate the specific case of ten particles, the first
beam comprises 3 particle and the second one 7. The particles move
on a straight line with the momenta $P_1$ and $P_2$, respectfully.
The states given by equation(\ref{ellipse}) are by no means equilibrium
thermodynamical states.

\noindent
Let us turn to the ideal gas in the Nos\'e --- Hoover thermostat.
On using momenta we may cast equations (\ref{ber}), (\ref{nos}) in the form

\begin{eqnarray}
  \dot{\vec p_i} &=& - \gamma \vec p_i   \label{nh} \\
  \dot{\gamma}   &=&  \alpha \left[ \frac{1}{3 m k_b T N}
				   \sum_{i=1}^N \dot{ \vec p_i}^2 - 1
			      \right]        \nonumber
\end{eqnarray}

\noindent
Consider again the two beams of $N_1$ and $N_2$ particles having initial
momenta $\vec P_1$ and $\vec P_2$, respectfully. Similarly to the situation
considered above, we may write equations (\ref{nh}) in the form of three
equations

\begin{eqnarray}
  \dot{\vec P_{\nu}} &=& - \gamma \vec P_{\nu}, \qquad \nu = 1,2
				  \label{nhred}                  \\
  \dot{\gamma}       &=&  \alpha
			      \left[ \displaystyle{\frac{1}{3 m k_b T N}}
				    (N_1 P_1^2 +  N_2 P_2^2) - 1
			      \right]        \nonumber
\end{eqnarray}

\noindent
Equations (\ref{nhred}) have stationary states given by the constraint
$$
  \gamma = 0
$$
and equation(\ref{ellipse}). In this sense the Nos\'e --- Hoover thermostat
is similar to the Berendsen one; they are also similar in that the
stationary states are not thermodynamical equilibria,
as is illustrated in Fig.2. But it is worth noting that Nos\'e --- Hoover's
thermostat, at least for the case of ideal gas we have been studying,
does not allow for attractor regimes, and in this sense it is even
farther from the real thermostat than Berendsen's one.

\noindent
{\bf 3. Conclusion}

\noindent
The example of ideal gas shows quite convincingly that neither
the Berendsen nor the Nos\'e --- Hoover thermostat can serve as a
device for modeling thermodynamical equilibrium. One may suggest,
of course, that for system more sophisticated than an ideal gas
they still could function to the effect, just owing to the
complexity of a system under consideration, certain models of macromolecules,
for example.

\noindent
All the same, it is worthwhile to note that
the Berendsen model isan  interesting example of
nonlinear dissipative problem, and its dynamics is characterized by
the presence of a set in its phase space
that attracts trajectories, that is an attractor.  The latter
is determined by the condition given by kinetic theory
$$
   \sum_{i=1}^N \, \displaystyle{\frac{\dot{\vec p_i}^2}{2 m}}
	     = \frac{3}{2} k_b T N
$$
In fact, it has also a more subtle structure of preserving the beams
of particles with equal momenta, as was discussed above.
Thus, at equilibrium the states of the system lie on a sphere in phase space.
If the equilibrium had been of thermodynamical nature, we should have had
a distribution of momenta corresponding to the canonical ensemble at a
temperature prescribed by the thermostat. Instead, we have the equation
(\ref{ellipse}), which describes the values of momenta for beams of
particles of the same momentum. Thus, there is no canonical ensemble in
the mechanical thermostat, which resembles more nonlinear dissipative
systems like the Brusselator, \cite{prig}. At any rate, the system arrives
at a stable state, following its evolution in time.

\noindent
In contrats, Nos\'e --- Hoover's system for ideal gas does not have
attractors, even though the set of its stationatry solutions is stable
(see Fig.2 and 3).   \vspace{2cm}

\centerline{FIGURE CAPTIONS  \vspace{5mm} }

\noindent
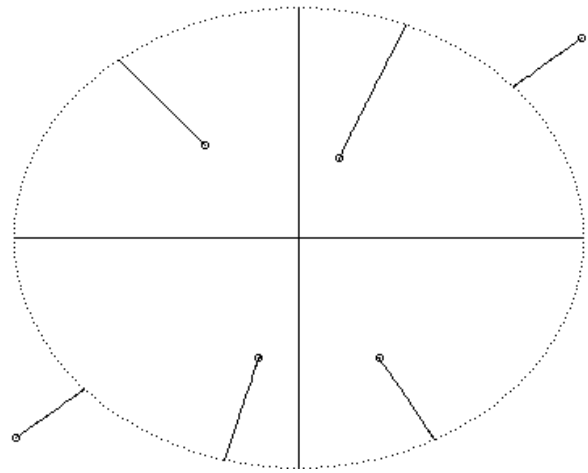
{\bf Fig.1} \\
Ten  one dimensional particles in the Berendsen thermostat.
The set of initial conditions for the momenta $P_1, P_2$ of the two
beams of 3 and 7 particles corresponds to the set of trajectories
which indicate the descent of the beams on the stationary states
belonging to the ellipse given by equation (\ref{ellipse}).

\noindent
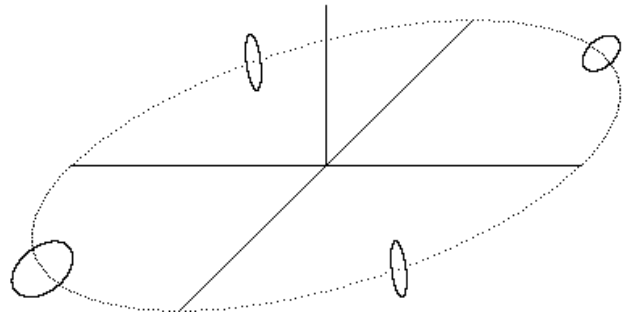
{\bf Fig.2} \\
Ten one dimensional particles in the Nos\'e --- Hoover thermostat.
The horizontal axes are $P_1, \, P_2$, the vertical one gives  values of
$\gamma$.  The closed curves indicate oscillatory motion round the set of
stationary states, which is stable. The partition of the system into
two beams is preserved, so that no thermodynamical equilibrium is
achieved.

\noindent
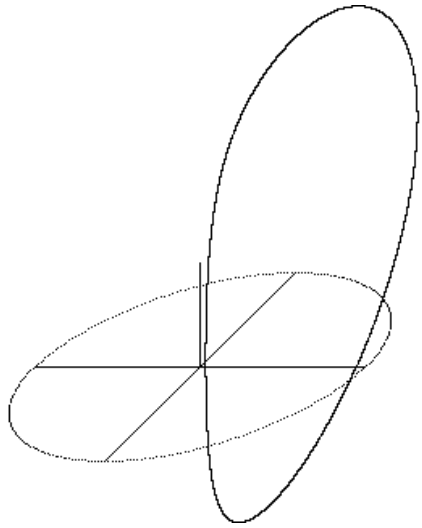
{\bf Fig.3} \\
Ten one dimensional particles in the Nos\'e --- Hoover thermostat.
The horizontal axes are $P_1, \, P_2$, the vertical one gives  values of
$\gamma$. A large amplitude motion of the beam;
no thermodynamical equilibrium.


\vspace{1cm}


\begin{thebibliography}{99}
	\bibitem{ber} H.J.C.Berendsen, J.P.M.Postma, and W.F. van Gunsteren,
	 A.DiNola, and J.R.Haak, Molecular dynamics with coupling to an
	 external bath, J.Chem.Phys. {\bf 81}, 3684 - 3690 (1984).
	\bibitem{nos} S.Nos\'e, Progr.Theor.Phys. Suppl. {\bf 103}, 1 - 46 (1991).
	\bibitem{rum} Yu.B. Rumer and M.Sh.Rivkin, Thermodynamics and Statistical
	 Physics, Nauka , Moscow (1977).
	\bibitem{prig} I.Prigogine and I.Stengers, La nouvelle alliance,
	 Gallimard, Paris (1979).
\end{thebibliography}
\end{document}